\documentclass[aps,prl,preprint,groupedaddress]{revtex4-1}

\usepackage{graphicx}
\begin{document}


\title{\Large  Conformal B-L  and Pseudo-Goldstone Dark Matter}


\author{\bf Rabindra N. Mohapatra$^a$}
\author{Nobuchika Okada$^b$}
\affiliation{}
\affiliation{$^a$ Maryland Center for Fundamental Physics and Department of Physics, University of Maryland, College Park, Maryland 20742, USA}
\affiliation{$^b$ Department of Physics, University of Alabama, Tuscaloosa, Alabama 35487, USA}


\date{\today}

\begin{abstract}
We show that a conformal extension of the standard model with local $B-L$ symmetry and two complex scalars breaking $B-L$ can provide a unified description of neutrino mass, origin of matter and dark matter.  
There are two  hierarchical $B-L$ breaking vacuum expectation value (VEV) scales in the model, the higher denoted by $v_B$ and the lower by $v_A$.  
The higher breaking scale is dynamically implemented  via the Coleman-Weinberg mechanism and plays a key role in the model since it induces electroweak symmetry breaking  as well as the lower $B-L$ breaking scale. 
It  is also responsible for neutrino masses via the seesaw mechanism and origin of matter. 
The imaginary part of the complex scalar with lower $B-L$ breaking VEV plays the role of  a pseudo-Goldstone dark matter (DM). 
The DM particle is unstable with its lifetime naturally longer than $10^{28}$ seconds. 
We show that its relic density arises from the freeze-in mechanism for a wide parameter domain. 
Due to the pseudo-Goldstone boson nature of the DM particle, the direct detection cross section is highly suppressed.  The model also predicts the dark matter to be heavier than 100 TeV and
it decays to two high energy neutrinos which can be observable at the  IceCube, providing a  test of this model. 


\end{abstract}

\maketitle

\section{1. Introduction} 
Understanding the origin of the electroweak scale is a fundamental puzzle of the standard model (SM). In usual discussion of the SM, it is customary to put the scale by hand in the form of a negative mass squared for the Higgs field.  
A more satisfactory approach, extensively discussed in the literature (see for some examples~\cite{C1, C1.1, C1.11, C1.2, C1.3, C1.4, C1.5, C1.6, C2, C3, C3.1, C4, C4.1, C4.2, C5, C5.2, C5.1, C5.3}) is  to consider conformal extensions of the SM where the mass terms vanish due to conformal symmetry and  the mass scales arise dynamically via the Coleman-Weinberg radiative correction mechanism~\cite{CW}.  
This approach when implemented in the SM leads to a very small mass for the Higgs boson and is ruled out by experiments. 
However, it can be implemented generally in the context of many extensions of the SM (see for instance ~\cite{C1,C1.1, C1.11, C1.2,C1.3,C1.4,C1.5,C1.6, C2, C3, C3.1, C4, C4.1, C4.2, C5,C5.2, C5.1, C5.3}). 
It is important to explore such models and pinpoint their tests. 
The goal of this paper is to present the phenomenological possibilities for dark matter in one such model. 

Local $B-L$ extension of the SM~\cite{marshak1,marshak2, davidson}  has been discussed as a very highly motivated minimal scenario for neutrino masses via the seesaw mechanism~\cite{ss1,ss2,ss3,ss4,ss5} and the origin of matter. 
Phenomenology of these models have been also extensively discussed in the literature; see for some examples \cite{BL1, BL2, BL3, BL3.1,BL3.2, BL4, BL5, MO1, BL5.1, BL6}. 
While most phenomenological discussion of the $B-L$ models have been done in connection with neutrino masses and collider physics, we recently pointed out another virtue \cite{MO1} of these models and showed that the real part of the SM singlet Higgs field that breaks $B-L$ can be a perfectly viable candidate for decaying dark matter. 
This dark matter model works only for very small values of the gauge coupling and low mass of the dark scalar (with masses in the MeV to a few keV range). The dark matter relic density in this case arises via the freeze-in mechanism \cite{hall} and has interesting experimental tests. 
The FASER experiment~\cite{faser} for example is ideally suited for testing this model.  
Here we present  a different approach  where we impose conformal invariance on this theory so that we can understand the origin of masses and explore if there is a dark matter candidate. 
We find that we need to extend the minimal $B-L$ model by enlarging the Higgs sector to include two complex SM singlet scalars which carry $B-L$ quantum numbers. 
In this case, the imaginary part of one of the two $B-L$ breaking scalars can be a viable dark matter candidate. 
In fact  the model predicts which particle can be the dark matter. 
It turns out that the dark matter is a psudo-Goldstone boson~\cite{pgb1, pgb2, pgb3, pgb4, pgb5, pgb6, pgb7, pgb8, pgb9, pgb10}, 
which explains why it  has escaped direct detection.

The two $B-L$ non-singlet scalars in our model have different VEVs. 
The higher scale $B-L$  breaking, which we call the primary breaking scale in the theory, arises from the Coleman-Weinberg mechanism by dimensional transmutation.  
The neutrino masses and the electroweak symmetry breaking~\cite{C2} as well as the lower scale $B-L$ breaking VEV  are all induced dynamically by the high scale $B-L$ breaking VEV.  
We show that in a wide parameter range, the imaginary part of the second $B-L$ breaking complex scalar is constrained to have a very long life~\cite{farinaldo, icecube, magic} and  plays the role of dark matter.  
An interesting prediction of the model is that  the dark matter mass is more than 100 TeV with its dominant decay mode being to two neutrinos. The model can therefore be tested by observation of TeV neutrinos by the ICE CUBE experiment. The baryon asymmetry of the universe in this model is generated by the leptogenesis mechanism~\cite{Fukugita-Yanagida}.

The paper is organized as follows: after briefly introducing the model and discussing $B-L$ symmetry breaking in sec.~2. 
We discuss the pseudo-Goldstone dark matter ($\sigma$) in sec.~3 and its lifetime in sec.~4. 
We  show that the lightest $B-L$ breaking scalar is naturally long lived enough for it to play the role of dark matter. 
In sec.~5, we show how the freeze-in mechanism determines the relic density of dark matter and the various constraints which imply on the parameter space of the model. 
In sec.~6, we discuss the constraints on the parameter space of the model and in sec. 7, we give some tests and comment on other aspects of the model, such as leptogenesis.  In sec.8 we summarize our results and conclude.

\section{2. Brief overview of the model} 
Our model is based on the local $U(1)_{B-L}$ extension of the SM with gauge quantum numbers of fermion under $U(1)_{B-L}$ defined by their baryon or lepton number. The full gauge group of the model is $SU(3)_c \times SU(2)_L\times U(1)_Y\times U(1)_{B-L}$, where $Y$ is the SM hypercharge. We need to add three right handed neutrinos (RHNs) with $B-L=-1$ to cancel the $B-L$ gauge anomaly. 
The RHNs being SM singlets do not contribute to SM anomalies. The electric charge formula in this case is same as in the SM. 

For the Higgs sector, in addition to the SM Higgs doublet $H$ which has zero $B-L$, we include two SM singlet Higgs fields:
$\Phi_B$ with $B-L=-2$ and $\Phi_A$ with $B-L=+6$.  
The tree level potential is given by
\begin{eqnarray}
V(H, \Phi_A, \Phi_B) &=& \lambda_H (H^\dagger H)^2+\lambda_ A(  \Phi^\dagger_A \Phi_A)^2 +\lambda_ B( \Phi^\dagger_B \Phi_B)^2  \nonumber \\
 && +   \lambda_{HA} (H^\dagger H)(\Phi^\dagger_A \Phi_A) + \lambda_{HB} (H^\dagger H)( \Phi^\dagger_B \Phi_B) 
  + \lambda_{mix} (\Phi^\dagger_A \Phi_A) (\Phi^\dagger_B \Phi_B)  \nonumber \\
   && -\lambda_{AB}\Phi_A\Phi^3_B+h.c. 
\label{eq:potential}   
\end{eqnarray}
Note that there are no mass parameters in the potential neither any mass terms in the  fermion sector of the Lagrangian making the theory conformal invariant. 

We break the $B-L$ symmetry in stages: the first stage is by giving VEV to the $B-L=2$ field $\Phi_B$, with $\langle \Phi_B \rangle=v_B/\sqrt{2}$; 
at the second stage, $\Phi_A$ acquires a lower scale VEV, $\langle \Phi_A \rangle =v_A/\sqrt{2}$. 
As we will see, $v_A$ is induced by the VEV $v_B$.  
We will show that $v_B \gg v_A$ naturally. 
The second VEV is induced by the high scale VEV via the $\lambda_{AB}$ term in the scalar potential. 
To induce the SM electroweak symmetry breaking, we choose the potential parameter $\lambda_{HB} < 0$ and adjust $~|\lambda_{HB}|v^2_B$ to be of order of the electroweak VEV squared~\cite{C2}. 
We discuss all these below.

\section{3. Symmetry breaking and pseudo-Goldstone dark matter} 
The first stage of the symmetry breaking is induced by the Coleman-Weinberg mechanism as follows: 
We write the one-loop potential involving
the $B-L$ breaking fields $\Phi_{A,B}$ near the $\Phi_B$ VEV as 
\begin{eqnarray}
V^{\rm 1-loop}~=~\frac{3g^4_{BL}}{16\pi^2}\left(36|\Phi_A|^2+4|\Phi_B|^2\right)^2\left[\ln\left(\frac{36|\Phi_A|^2+4|\Phi_B|^2}{\mu^2}\right)-\frac{5}{6}\right] .
\end{eqnarray}
Now if we choose $\lambda_B\sim g^4_{BL}$ and $\lambda_A \gg g^4_{BL}$, 
then the field $\Phi_B$ will acquire VEV ($\langle \Phi_B \rangle=v_B/\sqrt{2}$) while $\Phi_A$ will have zero VEV.  
Clearly now, $\lambda_{HB}<0$ will induce a negative mass term for the SM Higgs field $H$ and will generate it a VEV. 
The magnitude of this VEV is adjusted by choosing the value of  $\lambda_{HB}$ appropriately. 
Thus, although the magnitude of the weak scale is not explained, its origin is more meaningful than in the usual discussion. 
We also note that there is no lower bound on the value of $|\lambda_{HB}|$ since the coupling is not induced by the $B-L$ gauge boson loop diagrams, 
and hence $v_B$ is a free parameter of the model. 

To induce the $\Phi_A$ VEV, we minimize the effective potential below the $v_B$ scale. We write $\Phi_A=\frac{1}{\sqrt{2}}(\varphi_A+i\chi_A)$.
Assuming $\lambda_{mix}$ very small and neglecting quantum corrections to $\lambda_A$, 
the effective potential below the mass scale $v_B$ can be written as 
\begin{eqnarray}
V_{eff}~=~-\frac{\lambda_{AB}}{4}v^3_B\varphi_A+\frac{\lambda_{A}}{4}(\varphi^2_A+\chi^2_A)^2.
\end{eqnarray}
Minimizing this effective potential, we find that
\begin{eqnarray}
v_A\simeq \left(\frac{\lambda_{AB}}{4 \lambda_A}\right)^{1/3} v_B;\\\nonumber
m_{\varphi'_A}\simeq \sqrt{3\lambda_A} v_A;~~~ m_{\sigma}\simeq \sqrt{\lambda_A}v_A, 
\end{eqnarray}
where $\varphi'_A \simeq \varphi_A+\beta \varphi_B$ and $\sigma\simeq \chi_A+\alpha \chi_B$ are the mass eigenstates
with $\beta, \alpha$ being of order $\frac{v_A}{v_B} \ll 1$, obtained through the mixing with $\varphi_B$ and $\chi_B$
defined by $\Phi_B=\frac{1}{\sqrt{2}}(v_B+\varphi_B+i\chi_B)$. 
Note that $\sigma$ is a pseudo-Goldstone boson. 
To see the pseudo-Goldstone boson nature of $\sigma$, note that in the limit of $\lambda_{AB}=0$, the theory has global $U(1)\times U(1)$ symmetry and there are two massless Nambu-Goldstone bosons; however once the $\lambda_{AB}$ term is switched on, the symmetry reduces to only one, 
the $U(1)_{B-L}$ gauge symmetry and the $\sigma$ field picks up mass proportional to $\lambda_{AB}$. 
It has the lowest mass among the $B-L$ breaking scalar fields and  is a pseudo-Goldstone boson.

Now adjusting $\lambda_{AB}$ we can make the $v_A$ much lower than the primary $B-L$ breaking scale $v_B$. 
From the mass calculation, we see that $\sigma$ is the lighter of the two particles $\varphi_A$ and $\sigma$ and can be the dark matter of the universe as we see in detail in the next section. 
It is a pseudo-Goldstone dark matter \cite{pgb1,pgb2,pgb3,pgb4,pgb5,pgb6,pgb7, pgb8}. 
In this discussion we have neglected the mixing between the $\chi_A$ and $\chi_B$ which will occur when both $v_A$ and $v_B$ are nonzero. 
This mixing is proportional to $\alpha\sim \frac{v_A}{v_B}$, which we take into account in our discussion below. 
The linear combination of fields $\chi_B^\prime \simeq \chi_B-\alpha \chi_A$ becomes the longitudinal mode of the $B-L$ gauge boson $Z'$ and the orthogonal combination 
$\sigma \simeq \chi_A+\alpha \chi_B$ becomes the dark matter. 
This is an unstable  dark matter, and we study its detailed properties in the next section.

Note that  the coupling $fNN\Phi_B^\dagger$ gives mass to the right handed neutrinos of magnitude $M_N \sim f v_B$ implementing the seesaw mechanism for neutrinos. Unlike the model in \cite{MO1}, the real part of $\Phi_B$ field decays rapidly as the universe evolves. 
The out-of-equilibrium decay of RHNs in our model is responsible for leptogenesis.

\section{4. Dark matter lifetime} 
As noted above, the field $\sigma$ can play the role of dark matter of the model, if it satisfies the lifetime constraints.  
There are two lifetime constraints: one from the search for cosmic ray neutrinos from decaying dark matter with IceCube~\cite{icecube}.  
This puts a lower bound on $\tau_{DM} \gtrsim 10^{28}$ sec for then mass of decaying dark matter in the range of $10^4 < m_{DM}[{\rm GeV}] < 10^9$.  
A second limit comes from the Fermi-Lat  search for gamma rays from dwarf spheroidal galaxies~\cite{farinaldo}. 
There are also limits from deep gamma ray survey from Perseus Galaxy Cluster by MAGIC collaboration \cite{magic}. 
As we will see below, the first one is directly applicable to our case and not the others.

Note that $\sigma$ does not couple to SM fermions at the tree level. 
To estimate its lifetime, we first study its decay properties.  
In our model, it is natural that $M_{N} > m_{\sigma}$, since $M_N\sim fv_B$ and $m_{\sigma}\simeq \sqrt{\lambda_A} v_A$ with $v_A \ll v_B$.
The possible decay modes of $\sigma$ are as follows:
 \begin{figure}[tb]
  \centering
 \includegraphics[width=0.7\linewidth]{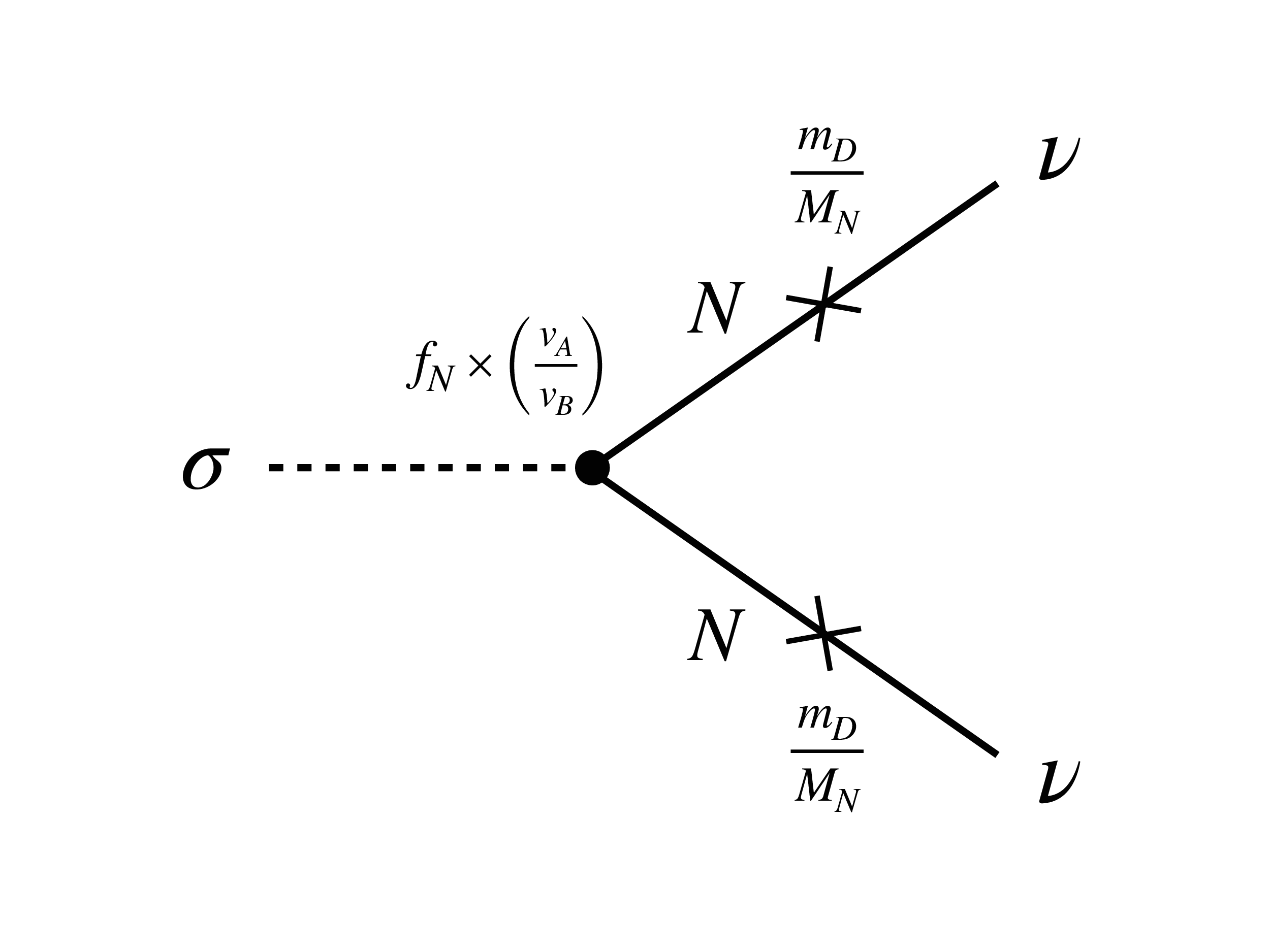} 
  \caption{Dominant decay mode of the dark matter $\sigma$ in the model to two neutrinos is shown. The decay proceeds through the mixing $\sim v_A/v_B$ of the dark matter with $\chi_B$ and the mixings of right handed neutrinos to light neutrinos in the seesaw mechanism $\sim m_D/M_N$.
} 
  \label{fig1}
  \end{figure}
(i)  $\chi_{\sigma}\to NN\to \nu \nu$ (see Fig.~\ref{fig1}).  
Here, the $N$ is a virtual state. 
The decay proceeds via the mixing of $\chi_{A}$ with $\chi_B$ proportional to $\alpha\sim \frac{v_A}{v_B}$ in amplitude. 
It is followed by the mixing between $N$ and $\nu$ through the seesaw mechanism $\sim m_D/M_N$, 
where $m_D$ is the neutrino Dirac mass. 
The decay width in this case can be estimated to be
\begin{eqnarray}
\Gamma_{\sigma \to \nu \nu} \simeq \frac{1}{4\pi}\left(\frac{m_\nu}{v_B}\right)^2\left(\frac{v_A}{v_B}\right)^2 m_\sigma. 
\label{eq:lifetime}
\end{eqnarray}
Here, we have used the seesaw formula for the light neutrino mass, $m_\nu \sim m_D^2/M_N$.  
%

(ii) Another possible mode is $\sigma\to \varphi_A^\prime+Z^\prime \to NN f\bar{f} \to \nu \nu f\bar{f}$, which is highly suppressed compared to 
$\Gamma_{\sigma \to \nu \nu}$ as is found to be
\begin{eqnarray}
\Gamma^{Z'}_{\sigma \to \nu\nu f \bar{f}}\simeq \frac{1}{ (4 \pi)^5} \left(\frac{m_\nu}{v_B}\right)^2\left(\frac{v_A}{v_B}\right)^2 \frac{m^5_{\sigma}}{v^4_B}.
\end{eqnarray}
The bottom line of the above discussion is that $\frac{\lambda_{AB}}{\lambda_A}\sim \left(\frac{v_A}{v_B}\right)^3$ must be very small for $\sigma$ to be a viable dark matter.
We also have assumed that $\lambda_{AH}$ is very small.

\section{5.  Dark matter relic density} 
In order to discuss how relic density of DM arises in this model, we use the freeze-in mechanism \cite{hall}. 
The requirement for the freeze-in mechanism is that the dark matter must be out of equilibrium with the cosmic soup of the SM particles. 
We first require that the reheat temperature of the universe after inflation $T_{R} < M_{Z^\prime}$. 
This is to avoid a resonance enhancement of the dark matter production by the $Z^\prime$ mediated process which leads to an over DM production unless the $B-L$ gauge coupling ($g_{BL}$) is extremely small. 

For $m_\sigma^2, m^2_{\varphi^\prime_A} \ll s \ll M_{Z^\prime}^2$, the DM annihilation process with $\varphi_A^\prime$ to a pair of SM fermions ($f \bar{f}$) is given by
\begin{eqnarray}
\sigma_{DM}\equiv\sigma_{{\sigma\varphi_A^\prime}\to f\bar{f}}\simeq \frac{g^4_{BL}s}{12\pi M^4_{Z'}} Q^2_{\Phi_A}(\sum_f N_f Q^2_f), 
\end{eqnarray}
where $(Q_f, N_f)=(1/3,3)$ for a quark, $(-1,1)$ for a charged lepton, and $(-1,1/2)$ for an SM neutrino, 
and $Q_{\Phi_A}=6$.
Counting all SM fermions and RHNs for the final states ($\sum_f N_f Q^2_f=8$) and using $M_{Z^\prime} \simeq 2 g_{BL} v_B$, 
this reduces to
\begin{eqnarray}
\sigma v_{rel}\simeq \frac{3 s}{\pi v^4_B}, 
\end{eqnarray}
where $v_{rel}$ is the relative velocity of the initial particles. 
For a temperature of the early universe $m_\sigma^2, m^2_{\varphi^\prime_A} \ll T^2 \ll M_{Z^\prime}^2$,
the thermal average of the above cross section is found to be
\begin{eqnarray}
   \langle \sigma v_{rel} \rangle \simeq \frac{36 T^2}{\pi v^4_B} = \frac{36 m_\sigma^2}{\pi v^4_B} x^{-2},
\label{eq:Xsec}
\end{eqnarray}
where $x \equiv \frac{m_\sigma}{T}$.


The freeze-in build-up happens via the reaction $f \bar{f} \to \sigma \varphi^\prime_A$ via $Z^\prime$ exchange and 
the DM yield $Y$ obeys the Boltzmann equation: 
\begin{eqnarray}
\frac{dY}{dx}\simeq \frac{ \langle \sigma v_{rel} \rangle}{x^2}\frac{s(m_\sigma)}{H(m_\sigma)} Y^2_{eq}, 
\label{Boltzmann}
\end{eqnarray}
where $Y_{eq} = \frac{n_{eq}(T)}{s(T)}$ with  the DM number density in thermal equilibrium $n_{eq}(T) \simeq \frac{T^3}{\pi^2}$, 
and the entropy density of the universe $s(T)=\frac{2 \pi^2 g_*}{45}T^3$,
and $H(T)=\sqrt{\frac{\pi^2}{90} g_*} \frac{T^2}{M_P}$ is the Hubble parameter 
with the reduced Planck mass $M_P=2.43 \times 10^{18}$ GeV  and  the effective degrees of freedom of the thermal plasma 
$g_*$  (we set $g_*=106.75$ in our analysis).  
We solve the Boltzmann equation from $x_R=m_\sigma/T_R \ll 1$ to $x=1$ with $Y(x_R)=0$. 
Note that for $T < m_\sigma \sim m_{\varphi_A^\prime}=\sqrt{3} m_\sigma$, or equivalently $x > 1$, the production of DM particles
is no longer effective since the averaged kinetic energy
of the SM particles in the thermal plasma becomes lower than $m_\sigma$.  
Using $\frac{s(m_{\sigma})}{H(m_{\sigma})}\simeq 13.7 \, m_{\sigma}M_P$, $Y_{eq} \simeq 2.16 \times 10^{-3}$ and Eq.~(\ref{eq:Xsec}), 
we get 
 \begin{eqnarray}
&&Y(x) \simeq  2.45 \times 10^{-4}  \, \frac{ M_P \, m_\sigma^3}{v_B^4} \left( \frac{1}{x_R^3} - \frac{1}{x^3}\right) \nonumber\\
& \to & Y(\infty) \simeq Y(x=1) \simeq 2.45 \times 10^{-4}  \, \frac{ M_P \, T_R^3}{v_B^4}. 
\label{eq:yield}
\end{eqnarray}
Note that the resultant $Y(\infty)$ value is determined by the reheating temperature, and this result is valid as long as
$1/x_R^3=(T_R/m_\sigma)^3 \gg 1$.  
We then use the formula for the current dark matter abundance to be
\begin{eqnarray}
\Omega_{DM}h^2 = \frac{m_\sigma s_0 Y(\infty)}{\rho_{crit} /h^2}, 
\label{eq:omega}
\end{eqnarray}
where $s_0\simeq  2890/{\rm cm}^3$ is the current entropy density, and $\rho_{crit}/h^2\simeq  1.05\times 10^{-5}$ GeV/cm$^3$. 
Using Eqs.~(\ref{eq:yield}) and (\ref{eq:omega}), we find
\begin{eqnarray}
T_R\simeq 9.02\times 10^{-9} \, v_B \left(\frac{v_B}{m_\sigma}\right)^{1/3} 
\label{eq:cond1}
\end{eqnarray}
to reproduce the observed DM relic abundance of $\Omega_{DM}h^2 =0.12$ \cite{Planck:2018vyg}. 

Let us now consider the out-of-equilibrium condition for the dark matter $\sigma$. 
The observed DM relic abundance $\Omega_{DM}h^2 =0.12$ leads to 
$Y(\infty) \simeq Y(1) \simeq \frac{4.36 \times 10^{-10}}{m_\sigma[{\rm GeV}]}$. 
Considering the fact that $Y(x)$ is a monotonically increasing function for $x_R \leq x \leq 1$ (see Eq.~(\ref{eq:yield})), 
we conclude that $Y(x) \leq Y(\infty) < Y_{eq}$ for $m_\sigma[{\rm GeV}] > 2.01 \times 10^{-7}$. 
This means that as long as $m_\sigma$ satisfies this lower bound, the yield $Y(x)$ starting from $Y(x_R)=0$ can never reach $Y_{eq}$
and therefore, the dark matter $\sigma$ has never been in thermal equilibrium.

\section{6. Summary of constraints on the model and parameter scan}

In this section, we summarize the constraints on the parameters of the model that add to the constraint in Eq.~(\ref{eq:cond1}) from the observed DM relic density.
The two main constraints that we have discussed are from the reheat temperature and the DM lifetime.

\noindent {\bf (A) Constraint from the freeze-in mechanism }

For the freezing mechanism to work, we need,
\begin{eqnarray}
   m_{\varphi^\prime_A, \sigma} < T_{R} <  M_{Z^\prime} .
\label{eq:cond2}   
\end{eqnarray}

\noindent {\bf (B) DM lifetime constraint:}
We next give the constraint from the dark matter life time:
\begin{eqnarray}
\Gamma_{\sigma \to \nu \nu} \simeq \frac{1}{4\pi}\left(\frac{m_\nu}{v_B}\right)^2\left(\frac{v_A}{v_B}\right)^2 m_\sigma
 <  6.58 \times 10^{-53} \; {\rm GeV},  
\label{eq:cond3} 
\end{eqnarray}
corresponding to the IceCube constraint \cite{icecube} on $\tau_\sigma > 10^{28}$ sec for $10^4 < m_\sigma[{\rm GeV}] < 10^9$. 
Note that this constraint implies $v_A \ll v_B$. 

\noindent {\bf (C) $\varphi_A^\prime$ lifetime constraint:} 

In addition to (A) and (B), we consider the constraint coming from the lifetime of $\varphi_A^\prime$. 
The main decay mode of $\varphi_A^\prime$ is $\varphi_A^\prime \to \sigma Z^\prime \to \sigma f \bar{f}$ via off-shell $Z^\prime$. 
This three-body decay width is calculated to be
\begin{eqnarray}
 \Gamma_{\varphi_A^\prime \to \sigma f \bar{f}} =\
 \frac{g_{BL}^4}{24 \pi^3} Q^2_{\Phi_A}(\sum_f N_f Q^2_f) \left(\frac{m_\sigma}{M_{Z^\prime}} \right)^4 m_{\varphi_A^\prime} C_I, 
\end{eqnarray}
where 
\begin{eqnarray}
C_I = \int_1^{\frac{m_{\varphi_A^\prime}^2+m_\sigma^2}{2 m_{\varphi_A^\prime} m_\sigma}} dz \,  (z^2-1)^{3/2} \simeq 0.0115,
\end{eqnarray}
for $m_{\varphi_A^\prime} = \sqrt{3} m_\sigma$. 
Note that $\Gamma_{\varphi_A^\prime \to \sigma f \bar{f}} \propto (v_A/v_B)^4$
from $M_Z^\prime \simeq 2 g_{BL} v_B$ and $m_\sigma \propto v_A$, and $\varphi_A^\prime$ can be long lived 
for $v_A \ll v_B$, which is required by the DM lifetime constraint. 
If $\varphi_A^\prime$ decays after the Big-Bang Nucleosynthesis (BBN), the energetic final state SM fermions
may destroy light nucleons successfully synthesized. 
To avoid this danger, we impose the following constraint that $\varphi_A^\prime$ decays before the BBN era,
at which the age of the universe is $\tau_{BBN} \simeq 1$ sec. 
\begin{eqnarray}
\Gamma_{\varphi_A^\prime \to \sigma f \bar{f}}
\simeq 4.84 \times 10^{-4} 
\left(\frac{m_\sigma}{v_B} \right)^4 m_\sigma > 6.58 \times 10^{-25} \; {\rm GeV}, 
\label{eq:cond4} 
\end{eqnarray}
where we have used 
$Q^2_{\Phi_A}=6$, $\sum_f N_f Q^2_f=8$, $M_{Z^\prime} \simeq 2 g_{BL} v_B$ and $m_{\varphi_A^\prime} =\sqrt{3} \, m_\sigma$
in evaluating $\Gamma_{\varphi_A^\prime \to \sigma f \bar{f}}$.

   \begin{figure}[tb]
  \centering
 \includegraphics[width=0.5\linewidth]{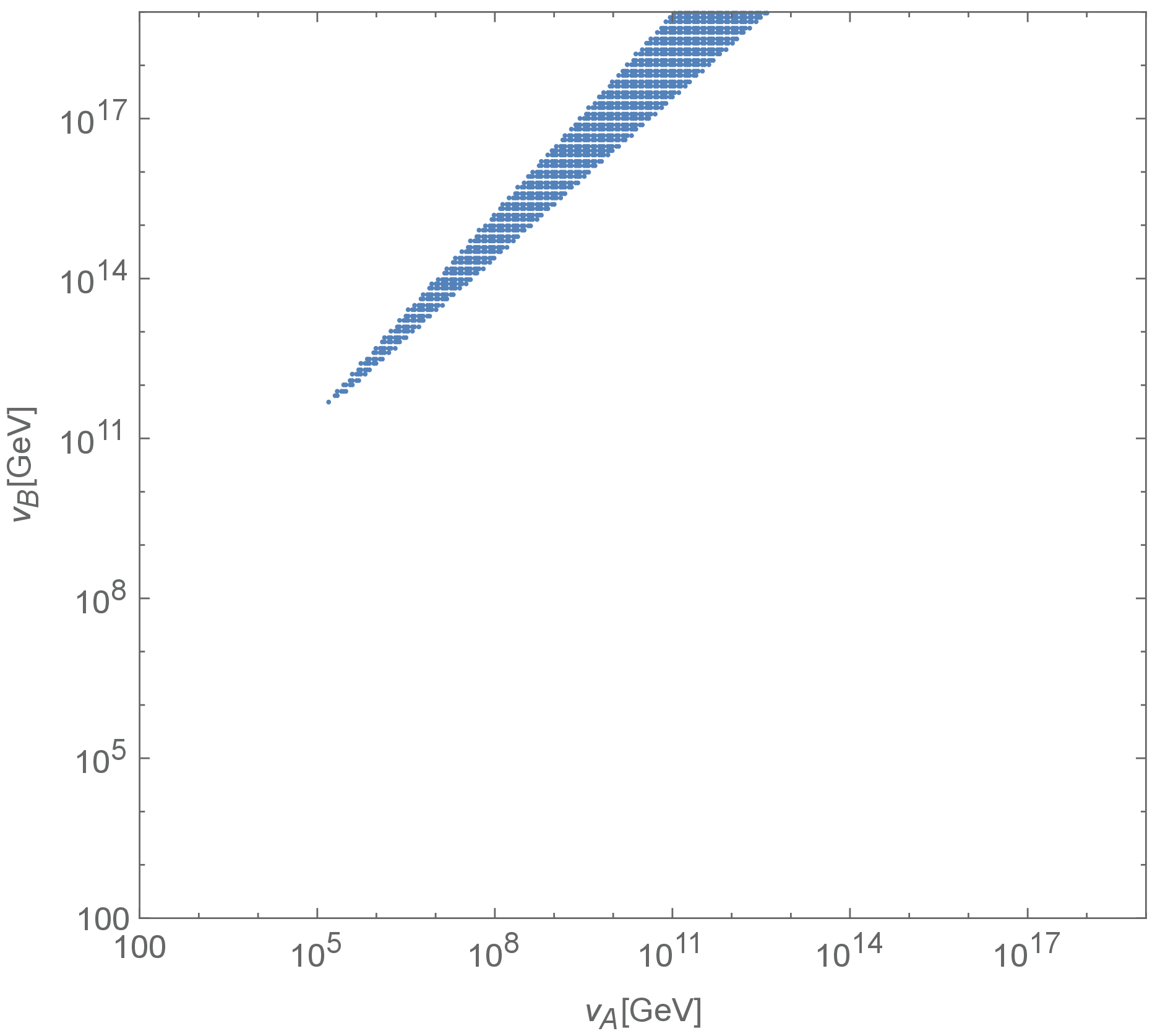}
  \caption{
  Allowed values of $v_A$ and $v_B$ parameters in the model are shown as the shaded region. 
} 
  \label{fig:plot1}
  \end{figure}

 \begin{figure}[tb]
  \centering
%
%
 \includegraphics[width=0.45\linewidth]{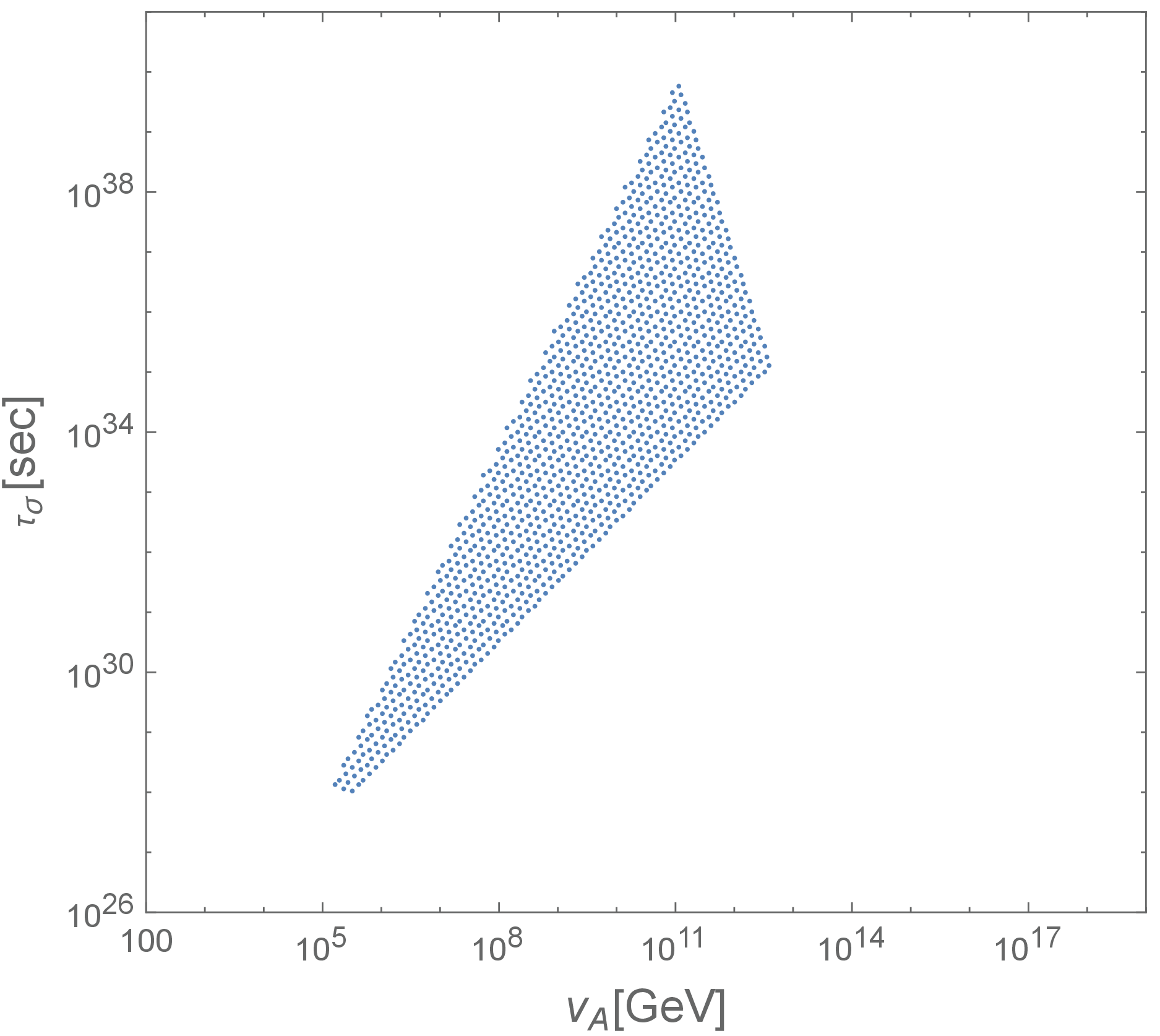}~~ 
 \includegraphics[width=0.45\linewidth]{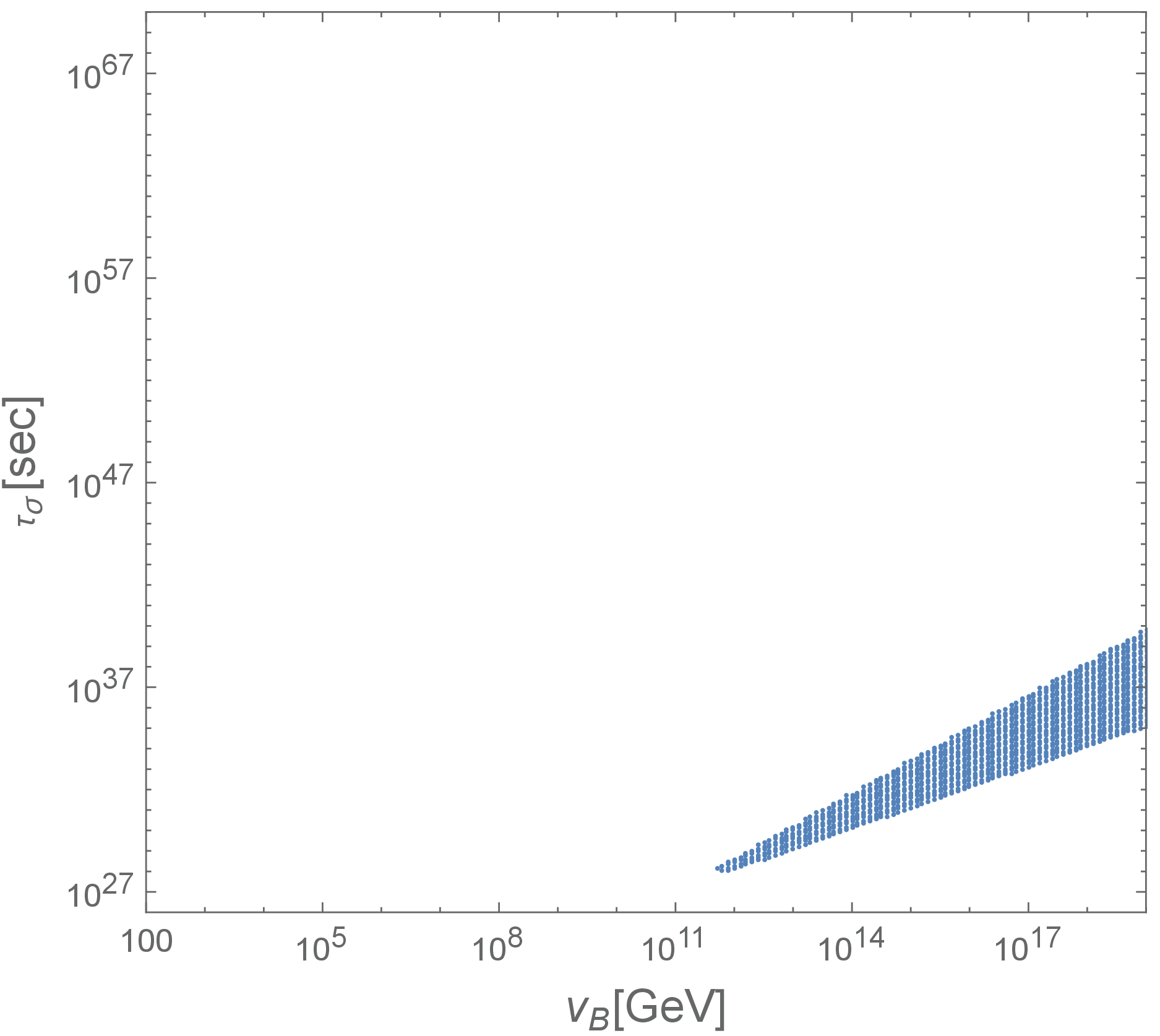} \\
  \caption{
Allowed range of the DM lifetime $\tau_\sigma$ and $v_{A, B}$. 
}  
  \label{fig:plot2}
  \end{figure}

 \begin{figure}[tb]
  \centering
%
%
 \includegraphics[width=0.45\linewidth]{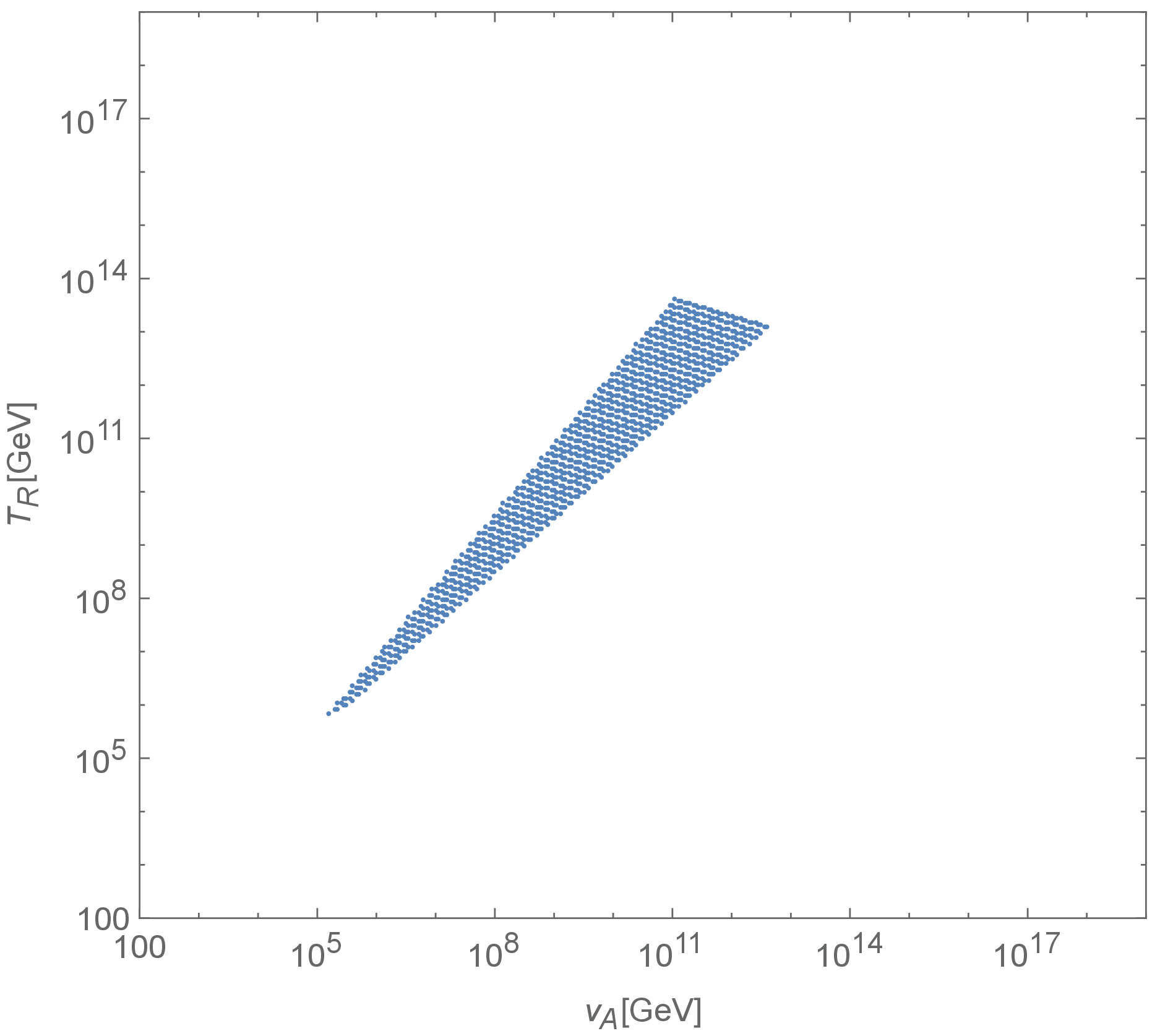}~~ 
 \includegraphics[width=0.45\linewidth]{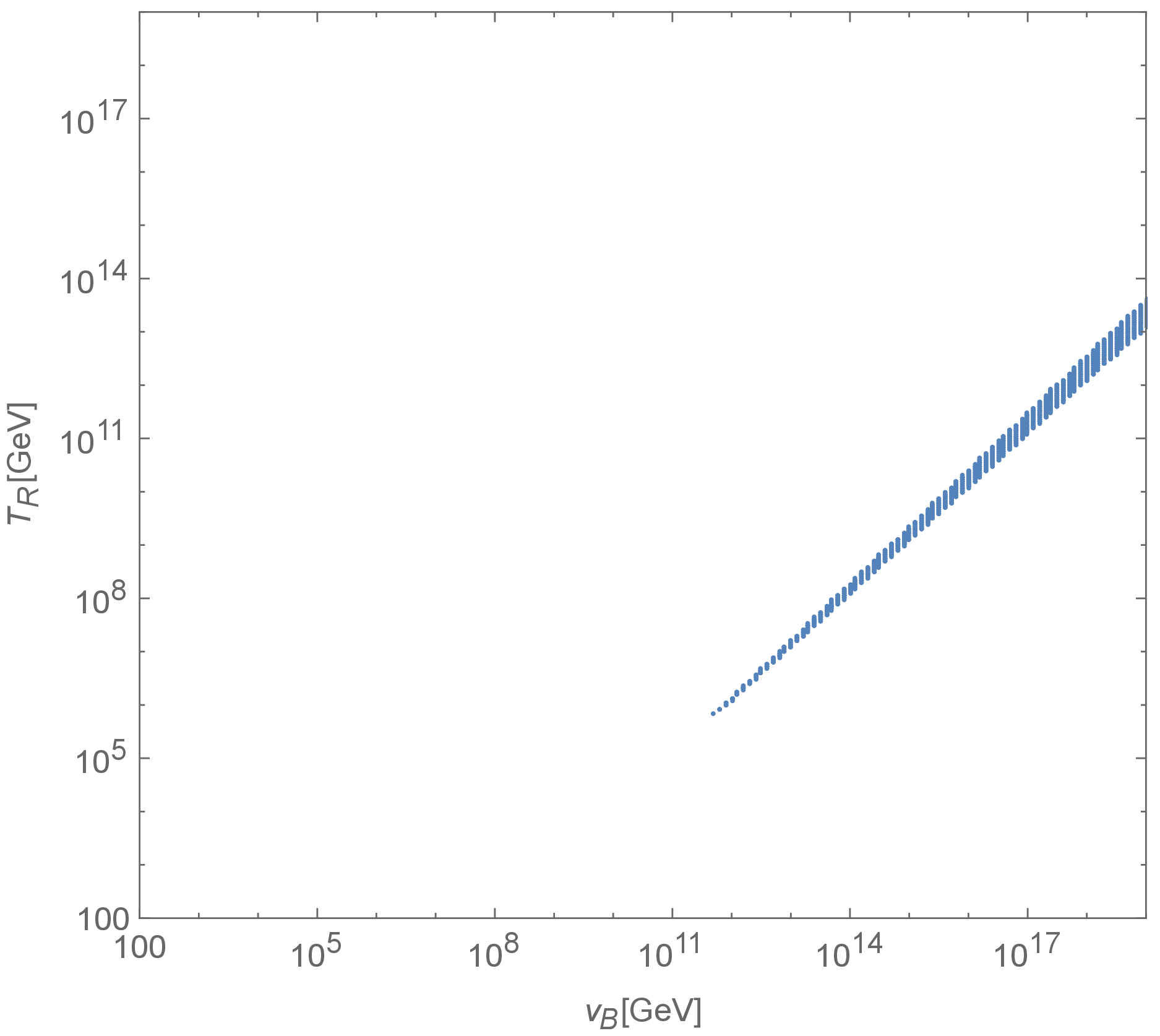} \\
  \caption{
Allowed range of the reheat temperature $T_R$ and $v_{A, B}$. 
}  
  \label{fig:plot3}
  \end{figure}

\begin{figure}[tb]
  \centering
  \includegraphics[width=0.5\linewidth]{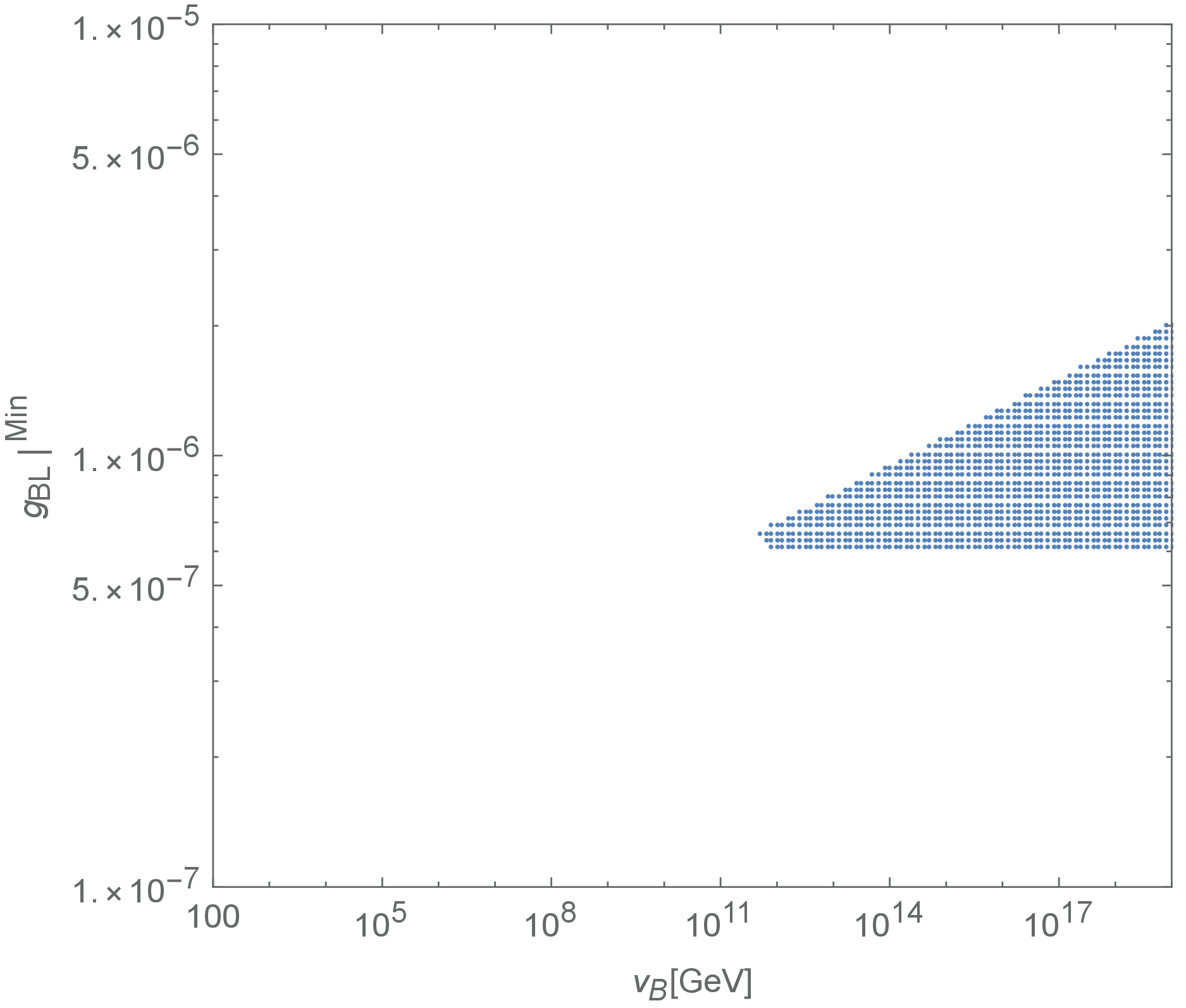}
  \caption{
The minimum values of the $g_{BL}$ coupling in the model for various values of $(v_A, v_B)$ chosen in the allowed region
in Fig.~\ref{fig:plot1}. 
} 
  \label{fig:plot4}
  \end{figure}

Imposing the conditions in Eqs.~(\ref{eq:cond1}), (\ref{eq:cond2}), (\ref{eq:cond3}) and(\ref{eq:cond4}), we perform the parameter scan
for two free parameters in the range of $100\, {\rm GeV} \leq v_A, v_B \leq 10^{19} \, {\rm GeV} $.
In this analysis, we set $m_\sigma = v_A$ for simplicity, and $T_R > 3 \, m_\sigma$ for the validity of Eq.~(\ref{eq:yield}). 
Fig.~\ref{fig:plot1} shows the allowed region in ($v_A, v_B$)-plane. 
Note that $m_\sigma=v_A \ll v_B$ is satisfied in the allowed region and the DM mass is constrained to be in the range of
$10^5 \, {\rm GeV} \lesssim m_\sigma \lesssim 10^{13}$ GeV. 
In Fig.~\ref{fig:plot2}, we show the range of the DM lifetime and $v_{A, B}$. 
We can see that the IceCube constraint on $\tau_\sigma > 10^{28}$ sec is most severe
for $m_\sigma \sim 10^5 \, {\rm GeV}$. 
Fig.~\ref{fig:plot3} shows the allowed range of the reheating temperature and $v_{A, B}$. 
The condition $T_R = M_{Z^\prime}$ sets the minimum value of gauge coupling 
$g_{BL}\left|^{\rm Min} \right.$ for a chosen values of $(v_A, v_B)$. 
For a set of parameters $(v_A, v_B)$ in the allowed region shown in Fig.~\ref{fig:plot1}, we show 
the output of $g_{BL}\left|^{\rm Min} \right.$ in Fig.~\ref{fig:plot4}. 
For the values of $(v_B, g_{BL}\left|^{\rm Min} \right.)$ selected in the shaded region, 
$g_{BL} > g_{BL} \left|^{\rm Min} \right.$ satisfies the condition $T_R < M_{Z^\prime}$. 


\section{7. Comments and phenomenology}
In this section we make a few comments on the model.

(i) We first note that the model  can explain the origin of matter via the usual leptogenesis mechanism. 
The temperature at which lepton asymmetry is generated depends on the particle spectrum of the model. 
For instance, note the important requirement of the model that $T_{R} <  M_{Z^\prime}$. 
So if $ v_{B}\sim 10^9$ GeV and $g_{BL}\lesssim 0.3 $ to avoid the coupling blowing up before the Planck scale, 
then the RHN mass must be less than $T_{R}$, which is too light to generate the observed baryon asymmetry by the usual leptogenesis. 
Thus, our scenario must use resonant leptogenesis mechanism \cite{res1, res2}.

(ii)  
Through the $B-L$ gauge interaction, RHNs can stay in thermal equilibrium with the SM particle plasma. 
As a result, the generation of lepton asymmetry is suppressed until the $B-L$ interaction is frozen. 
To avoid the suppression, we impose $n_{eq} \langle \sigma v_{rel} \rangle < H$ at $T \sim M_N$, 
where $n_{eq} \simeq 2 M_N^3/\pi^2$ is the RHN number density, and $\langle \sigma v_{rel} \rangle$ 
is the thermal averaged cross section for the process $NN \leftrightarrow f \bar{f}$ via a virtual $Z^\prime$, roughly given by
\begin{eqnarray}
  \langle \sigma v_{rel} \rangle \simeq \frac{g_{BL}^4}{4 \pi} \frac{M_N^2}{M_{Z^\prime}^4} \simeq 
   \frac{M_N^2}{64 \pi \, v_B^4}. 
   \label{eq:XsecN}
\end{eqnarray}
We then find the condition, 
\begin{eqnarray}
  M_N < (32 \pi^3)^{1/3} \left(\frac{\pi^2}{90} g_* \right)^{1/6} v_B \left(\frac{v_B}{M_P} \right)^{1/3}
  \simeq 10.3 \times v_B \left(\frac{v_B}{M_P} \right)^{1/3}. 
  \label{eq:cond5} 
\end{eqnarray}
In Fig.~\ref{fig:plot1}, we have found the lower bound on $v_B \gtrsim 10^{11}$ GeV, for which the condition reads $M_N \lesssim 3.54 \times 10^9$ GeV. 
For the successful (resonant) leptogenesis scenario, the lightest RHN must be in thermal equilibrium for $T_R \geq T > M_N$. 
Our result shown in Fig.~\ref{fig:plot3} indicates that the condition of Eq.~(\ref{eq:cond4}) is satisfied for the resultant $T_R$. 

If $M_N > m_{\varphi_B^\prime}$, the generation of lepton asymmetry can be suppressed by the process,
$NN \leftrightarrow \varphi_B^\prime \varphi_B^\prime$ \cite{Dev:2017xry}. 
To avoid this suppression, we impose this process to be decoupled at $T \sim M_N$. 
The thermal averaged cross section for this process is given by 
\begin{eqnarray}
  \langle \sigma v_{rel} \rangle \simeq \frac{f^4}{4 \pi \, M_N^2} \simeq \frac{M_N^2}{4 \pi \, v_B^4}. 
\end{eqnarray}
This formula is the same as Eq.~(\ref{eq:XsecN}) up to a factor, so that the resultant constraint is similar to Eq.~(\ref{eq:cond5}). 
The similar processes, $NN \leftrightarrow \varphi_A^\prime \varphi_A^\prime$, $\sigma \sigma$, have no effect, 
since they are suppressed by the mixing $v_A/v_B \ll1$ and $\phi_A^\prime$ and $\sigma$ are out-of-equilibrium. 



(iii) One possible test of the model is that the dark matter decays into two neutrinos with energy $E_\nu= \frac{m_{\sigma}}{2}$ for each neutrinos and for dark matter masses in the multi-TeV range, there will be high energy mono-energetic neutrinos from the DM decay with a probability of $\tau_{U}/\tau_\sigma$, where $\tau_U \sim 10^{17}$ sec is the age of the universe.

(iv) We also note that due to pseudo-Goldstone nature of the DM,  direct detection cross section arises only at one loop level and is highly suppressed. At the tree level the DM behaves like an inelastic dark matter since $Z'$ exchange by incident DM connects to a $\varphi_A^\prime$ field which is $\sqrt{3}$ times heavier. This explains why the DM has not been seen in the laboratory experiments.

(v) This model can be extended to allow the real part of the $\Phi_B$ field to play the role of inflaton while maintaining conformal invariance, as has been shown in Ref.~\cite{Nobuinf, Nobuinf2, Nobuinf3}.  
In this case as well as in general, there is an upper limit on the reheat temperature coming from the power spectrum and upper limit on the tensor-to-scalar ratio $r \leq 0.036$ at 95\% confidence level \cite{BICEP:2021xfz}.
Typically, 
$T_R \gtrsim 6 \times 10^{15}$ GeV is ruled out, assuming the total inflaton energy is transmitted to the SM thermal plasma 
right after inflation.  

\section{8. Summary}
In this brief note, we have presented a minimal conformal $B-L$ extension of the standard model which explains the neutrino masses, origin of matter and a dark matter that is produced  in the early universe by the freeze-in mechanism. 
We have presented the allowed set of points where the model works.  We find that the dark matter must be heavier than 100 TeV in order to ensure that its partner $\varphi'_A$ must decay before the big bang nucleosynthesis.
The model predicts energetic neutrinos from dark matter decay (with $E_\nu \geq 50$ TeV)  which can be observed at the IceCube experiment, providing a test. 

\section*{Acknowledgement}
The work of N.O. is supported in part by the US Department of Energy grant no.~DE- SC0012447.


\end{document}